\documentclass[12pt]{article}

\usepackage{amsmath,amssymb,amsfonts,amsbsy}
\usepackage{graphicx}
\usepackage{wrapfig}

\begin{document}
\addcontentsline{toc}{subsection}{{Title of the article}\\
{\it L.V. Nogach}}

\setcounter{section}{0}
\setcounter{subsection}{0}
\setcounter{equation}{0}
\setcounter{figure}{0}
\setcounter{footnote}{0}
\setcounter{table}{0}

\begin{center}
\textbf{  A FEASIBILITY EXPERIMENT AT RHIC TO MEASURE \\ THE ANALYZING 
POWER FOR DRELL-YAN PRODUCTION (A$_N$DY)}

\vspace{5mm}

\underline{L.~Nogach}$^{\,4,\dag}$, E.C.~Aschenauer$^{\,1}$, A.~Bazilevsky$^{\,1}$, 
L.C.~Bland$^{\,1}$, K.~Drees$^{\,1}$, C.~Folz$^{\,1}$, Y.~Makdisi$^{\,1}$, 
A.~Ogawa$^{\,1}$, P.~Pile$^{\,1}$, T.G.~Throwe$^{\,1}$, H.J.~Crawford$^{\,2}$, 
J.~Engelage$^{\,2}$, E.G.~Judd$^{\,2}$, C.W.~Perkins$^{\,2,\,3}$,
A.~Derevshchikov$^{\,4}$, N.~Minaev$^{\,4}$, D.~Morozov$^{\,4}$,
G.~Igo$^{\,5}$, M.~Grosse Perdekamp$^{\,6}$, M.X.~Liu$^{\,7}$, H.~Avakian$^{\,8}$, 
E.J.~Brash$^{\,8,\,9}$, C.F.~Perdrisat$^{\,10}$, V.~Punjabi$^{\,11}$, X.~Li$^{\,12}$, 
M.~Planinic$^{\,13}$, G.~Simatovic$^{\,13}$, A.~Vossen$^{\,14}$, 
G.~Schnell$^{\,15}$, C.~Van~Hulse$^{\,15}$, A.~Shahinyan$^{\,16}$, S.~Abrahamyan$^{\,16}$, 
N.~Liyanage$^{\,17}$, K.~Gnanvo$^{\,17}$

\vspace{5mm}

\begin{small}
  (1) \emph{Brookhaven National Laboratory, USA} \\
  (2) \emph{University of California/Space Science Laboratory, Berkeley, USA} \\
  (3) \emph{Stony Brook University, USA} \\
  (4) \emph{Institute of High Energy Physics, Protvino, Russia} \\
  (5) \emph{University of California, Los Angeles, USA} \\
  (6) \emph{University of Illinois, USA} \\
  (7) \emph{Los Alamos National Laboratory, USA} \\
  (8) \emph{Thomas Jefferson National Accelerator Facility, USA} \\
  (9) \emph{Christopher Newport University, USA} \\
  (10) \emph{College of William and Mary, USA} \\
  (11) \emph{Norfolk State University, USA} \\
  (12) \emph{Shandong University, China} \\
  (13) \emph{University of Zagreb, Croatia} \\
  (14) \emph{Indiana University, USA} \\
  (15) \emph{University of the Basque Country and IKERBASQUE, Spain} \\
  (16) \emph{Yerevan Physics Institute, Armenia} \\
  (17) \emph{University of Virginia, USA} \\

  $\dag$ \emph{E-mail: Larisa.Nogach@ihep.ru}
\end{small}
\end{center}

\vspace{0.0mm} 

\begin{abstract}
Large transverse single spin asymmetries (SSA) were measured for
pions produced in $p$$\uparrow$$p$-collisions up to RHIC energies. Sizeable
SSA were also found in semi-inclusive deep inelastic scattering (SIDIS).
Theory can explain such spin effects by going beyond collinear leading-twist 
perturbative QCD (pQCD) to include transverse momentum dependent (TMD) 
distribution and fragmentation functions. One of the most interesting 
TMDs is the Sivers function, which provides information on the correlation 
between the transverse spin of the nucleon and the transverse momentum 
distributions of the partons in the nucleon. It is particularly intriguing 
that theory predicts the Sivers function will change sign from SIDIS to 
Drell-Yan (DY) production. A$_N$DY is aiming to test that prediction and 
to establish requirements for future upgrades at RHIC to study DY production. 
The experiment configuration, achievements to date, status and plans are discussed.
\end{abstract}

\vspace{7.2mm} 

\section{A feasibility experiment to measure Drell-Yan}
\vspace*{-1mm} 

RHIC remains a unique machine that can accelerate and collide  polarized 
proton beams at center-of-mass-energies $62 \le \sqrt{s} \le 500~GeV$. The goal 
of the RHIC spin program is to identify how the proton gets its spin from the 
quark and gluon constituents. Transverse SSA play an important role in 
understanding the spin structure of the proton because in the simple picture 
of collinear leading-twist perturbative QCD, they are expected to be small. 
Contrary to these expectations, significant spin effects in inclusive pion 
production in $pp$-collisions were first found at low energies ($\sqrt{s} \le 20~GeV$) 
\cite{ln_e704,ln_e925} and then measured at RHIC~\cite{ln_fpd,ln_brahms}. 
Non-zero SSA were also observed in SIDIS from transversely polarized proton 
targets~\cite{ln_hermes,ln_compass}. The experimental results have stimulated 
theory development which led to extensions of the collinear parton model by 
introducing spin-correlated transverse momentum ($k_T$) to parton distribution 
and fragmentation functions. 

The Sivers mechanism attributes the transverse spin effects to a correlation 
between the parton $k_T$ and spin of the proton~\cite{ln_sivers}. To give a 
non-zero effect, it requires a final-state interaction in SIDIS. Theory 
predicts that the attractive final-state interaction in SIDIS becomes 
a repulsive initial-state interaction in DY process thereby resulting in 
a sign change for the Sivers function between the two processes~\cite{ln_collins}. 
New theoretical development and attempts of a global analysis of SIDIS and inclusive 
pion production at RHIC \cite{ln_kang,ln_ap} lead to the conclusion that it is
essential to test the predicted sign change for DY in the $x_F$ region that overlaps 
with SIDIS kinematics. 

\begin{wrapfigure}[17]{R}{70mm}
  \centering 
  \vspace*{-1mm} 
  \includegraphics[trim = 0mm 40mm 0mm 10mm, clip, width=70mm]{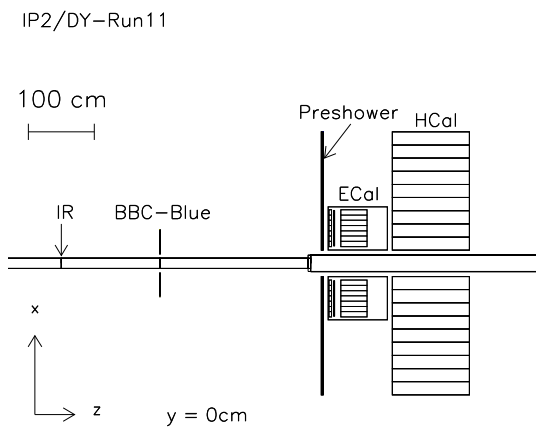}
  \caption{\footnotesize
Top view of A$_N$DY configuration for RHIC 2011 run. The Blue beam travels in 
the positive z direction, and the Yellow beam in the opposite direction. 
IR indicates the center of the collision region.
}
  \label{ln_fig1}
\end{wrapfigure}
A feasibility experiment at RHIC, A$_N$DY, was proposed to test that 
prediction and to establish basic requirements for future upgrades at RHIC 
for DY measurements. Forward DY production is of interest not just for 
the analyzing power, it is also the most robust observable sensitive to 
low-$x$ parton distributions for intercomparison to results in $dAu$ pion 
production and to a future electron-ion collider. A$_N$DY is located at RHIC 
2 o'clock Interaction Point (IP2), where beam polarization is always transverse. 
It thus can run in parallel with the RHIC $W$ program at IP6 and IP8. A$_N$DY goals 
are: 1) to establish that large-$x_F$ low-mass $e^+e^-$-pairs from the DY process 
can be discriminated from background in $pp$-collisions at $\sqrt{s}=500~GeV$; 
2) to provide sufficient statistical precision for the DY analyzing power measurement 
to test the theoretical prediction of a sign change compared to SSA for SIDIS. Yet 
another objective is to determine whether tracking for the charge sign discrimination 
is necessary for DY measurements or calorimetry alone would be sufficient. 

\vspace*{-4mm} 
\section{2011 test run}
\vspace*{-1mm} 

A schematic view of the A$_N$DY setup for the 2011 run is shown in Fig.~\ref{ln_fig1}. 
The setup included: two beam-beam counters (BBC), located on both sides of the IR 
(BBC-Yellow is not shown), for minimum-bias triggering and luminosity measurement; 
two zero-degree calorimeters with shower maximum detectors (ZDC, not shown) for 
luminosity measurement and verifying that A$_N$DY can measure a known analyzing 
power; hadron calorimeter (HCal)~--- two modules of 9$\times$12 lead-scintillating 
fiber cells placed symmetrically left and right of the beam pipe, to assess hadronic
background and for jet reconstruction; small electromagnetic calorimeter 
(ECal)~--- two 7$\times$7 matrices of (4~cm)$^2\times$40~cm lead glass cells,
for photon/electron/$\pi^0$-reconstruction; preshower detector (two planes, 2.5~cm 
and 10~cm scintillating strips) to assist in photon/electron/hadron separation.

\begin{wrapfigure}[15]{R}{75mm}
  \centering 
  \vspace*{-5mm} 
  \includegraphics[trim = 0mm 5mm 0mm 0mm, clip, width=75mm]{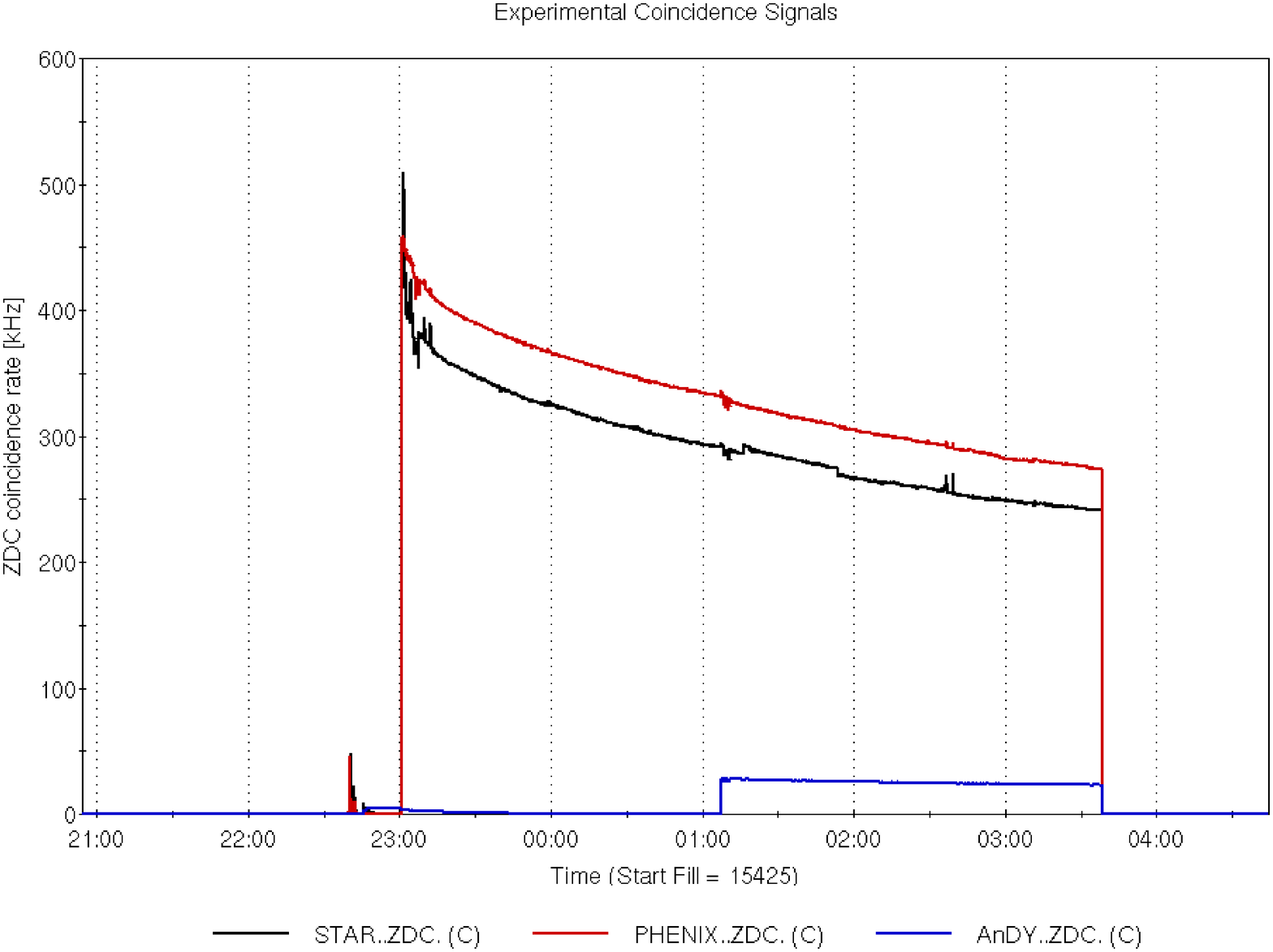}
  \caption{\footnotesize
ZDC coincidence rate at three interaction points versus time for one RHIC store. 
Collisions at IP2 were initiated after bunch intensity had decreased to 
1.5$\times$10$^{11}$ ions per bunch.
}
  \label{ln_fig2}
\end{wrapfigure}
The primary goals for the 2011 run were:\\ 1) to establish the impact of collisions
at IP2 on IP6 (STAR) and IP8 (PHENIX) operation; 2) to demonstrate a means of 
HCal calibration; 3) to measure the hadronic background for comparison with 
simulations. Fig.~\ref{ln_fig2} shows luminosities at three interaction points 
for a RHIC fill. Collisions at IP2 began when the bunch intensity had decreased
to a preset value (in this case, 1.5$\times$10$^{11}$ ions per bunch).     
There is no noticeable effect from adding collisions at IP2 for this fill.
The Collider-Accelerator Department (CAD) developed an automated procedure for
bringing A$_N$DY into collisions, and repeatedly demonstrated that it can be 
done without significant impact on beam life time and luminosities at the other
two interaction points. Along with these tests, A$_N$DY was taking data using
a set of triggers: minimum bias (a condition on the time difference between 
the earliest hits in the BBC-Yellow and BBC-Blue to define a collision at IP2), 
energy sum in ECal, jets in HCal for physics, LED and cosmic-rays for monitoring. 
Integrated luminosity of 6.5~$pb^{-1}$ was acquired. DY measurements set the goal 
of $\ge 100~pb^{-1}$. CAD stated that $\sim$~10~$pb^{-1}$/week could be delivered 
to IP2, but would require decreasing $\beta^*$ from 3~m to 1.5~m. With 100~$pb^{-1}$ 
of integrated luminosity and 50\% beam polarization expected uncertainty for 
the DY analyzing power measurements is $\sim$~0.03.

\begin{figure}[b!]
  \centering
  \begin{tabular}{ccc}
    \includegraphics[width=50mm]{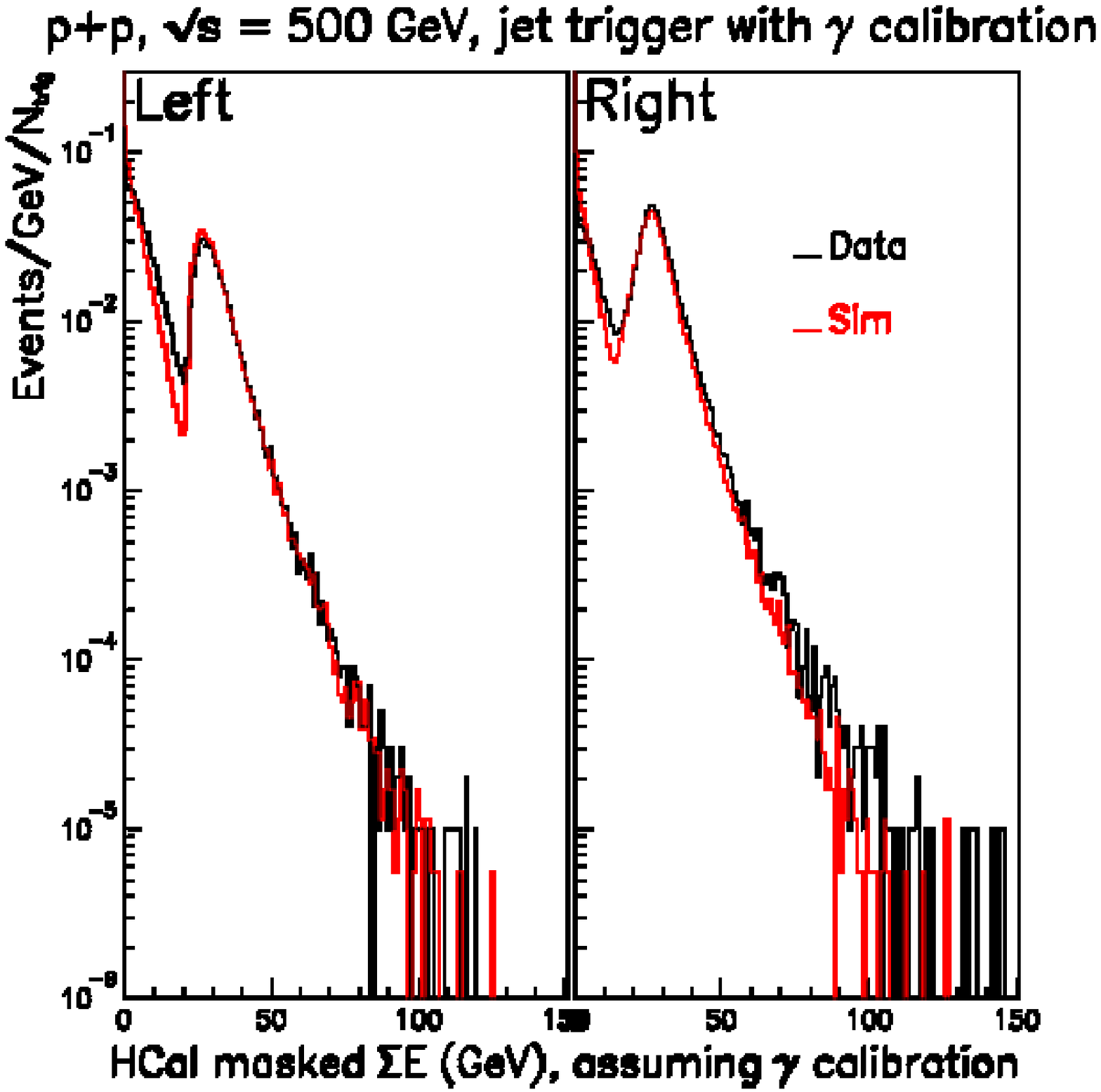} &
    \includegraphics[width=50mm]{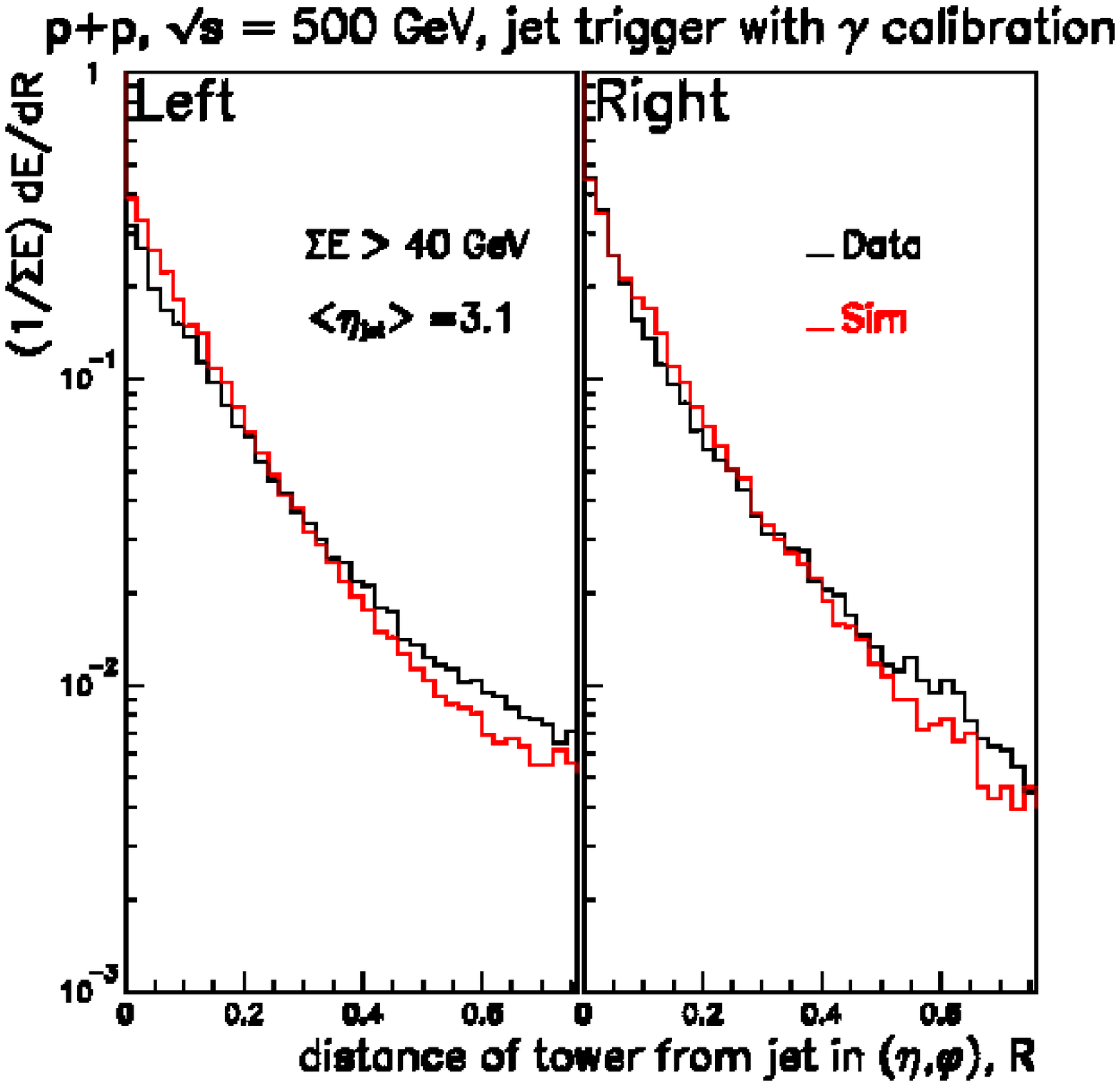} &
    \includegraphics[width=50mm]{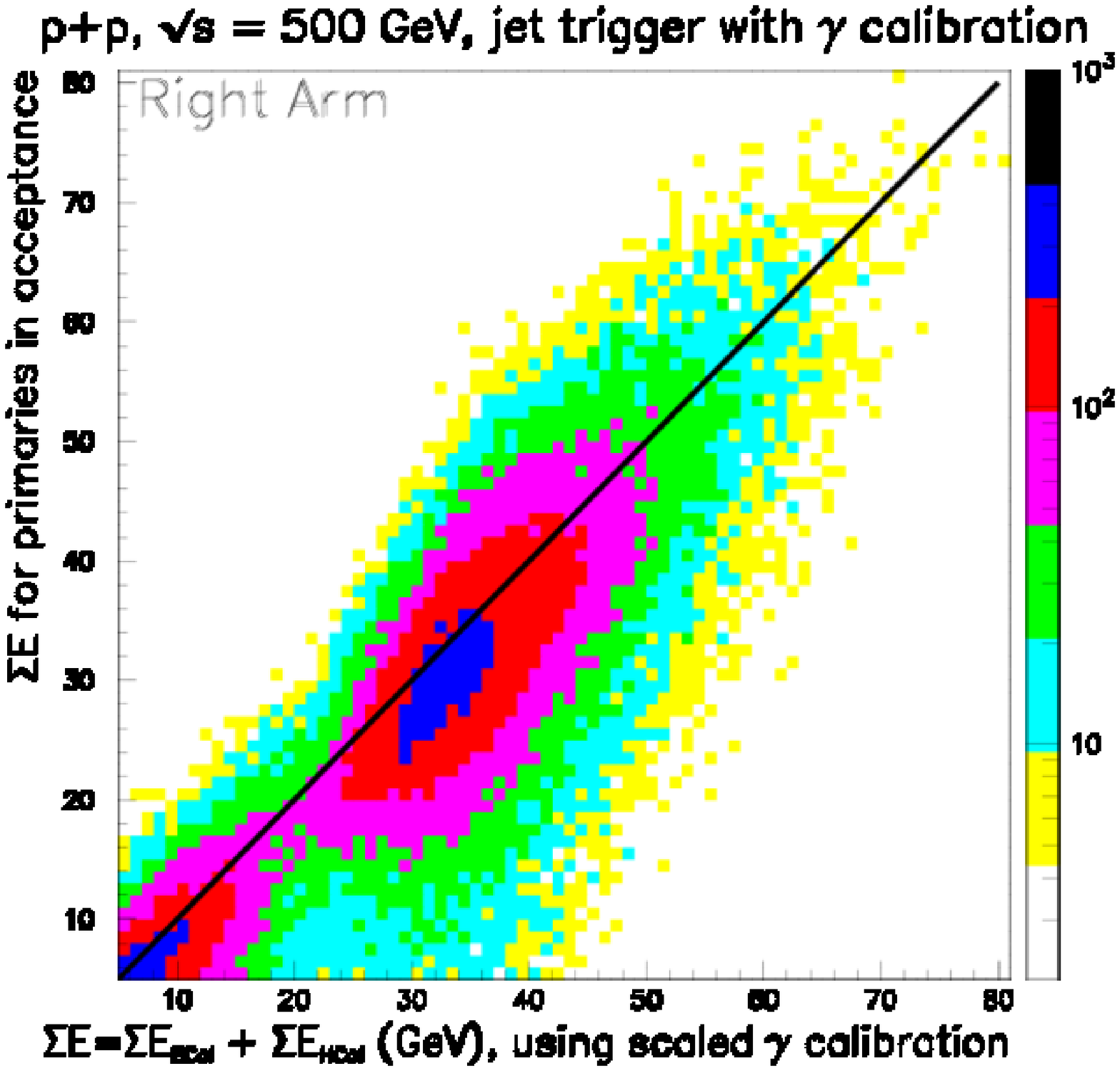} \\
    \textbf{(a)} & \textbf{(b)} & \textbf{(c)}
  \end{tabular}
  \caption{\footnotesize
    \textbf{(a)} Distribution of summed energy in HCal modules from the jet trigger.
    \textbf{(b)} Distribution of energy in the jet as a function of distance 
in $\eta$-$\phi$ space from the jet center (jet shape).
    \textbf{(c)} Jet energy from simulations versus reconstructed response
in ECal+HCal.
  }
  \label{ln_fig3}
\end{figure}
About 7.5$\times 10^8$ jet triggered events were collected during 2011 run. First 
step in the data analysis was to define the absolute energy scale of HCal. This 
was done using $\pi^0$-reconstruction~\cite{ln_cp_dis2011}. An attempt at full 
jet reconstruction using this calibration is shown in Fig.~\ref{ln_fig3}. There is 
a reasonable agreement between data and PYTHIA (version 6.222) + GEANT simulations 
for jet energy (Fig.~\ref{ln_fig3}a). The data show a somewhat broader jet shape 
than the simulations (Fig.~\ref{ln_fig3}b). Flattening of the jet shape at larger R 
is caused by an underlying event contribution, which appears to be smaller than a few 
percents. The jet energy scale was checked with the association analysis of the 
simulations: primaries from PYTHIA were projected to HCal, the energy of those 
in the acceptance was summed and correlated with the reconstructed energy in ECal+HCal. 
This correlation is shown in Fig.~\ref{ln_fig3}c, and it proves that the jet energy 
scale is quite well defined using the photon (neutral pion) calibration. The next step 
would be the more sophisticated algorithm for jet reconstruction and a look at 
the jet analyzing power. 

\vspace*{-4mm} 
\section{Future plans}
\vspace*{-1mm} 

The A$_N$DY proposal that was endorsed by the Program Advisory Committee at BNL 
in May 2011, supposed two years of data taking for DY measurements: a 2012 run with 
the full calorimetry and a 2013 run with the magnet and tracking added. The initial 
approach was to build modular, left-right symmetric detectors for A$_N$DY. To reduce 
the luminosity requirements and taking into account that for the most robust 
measurement of the sign change the $x_F$ region would overlap with SIDIS kinematics, 
the A$_N$DY acceptance was optimized. ECal, HCal and preshower detector will be stacked 
around the beam pipe to provide full azimuthal coverage. Electromagnetic calorimetry 
required for the complete A$_N$DY setup was loaned from JLab and delivered to BNL. 
The staging of the apparatus awaits a funding review. The first attempt at a transverse 
spin DY measurement will be in the RHIC 2013 run.


\begin{thebibliography}{99} 

\bibitem{ln_e704}
D.L.~Adams \textit{et al.}, Phys. Lett. \textbf{B261} (1991) 201; 
\textbf{B264} (1991) 462.

\bibitem{ln_e925}
K.~Krueger \textit{et al.}, Phys. Lett. \textbf{B459} (1999) 412; \\
C.E.~Allgower \textit{et al.}, Phys. Rev. \textbf{D65} (2002) 092008. 

\bibitem{ln_fpd}
B.~Abelev \textit{et al.}, Phys. Rev. Lett. \textbf{101} (2008) 222001.

\bibitem{ln_brahms}
I.~Arsene \textit{et al.}, Phys. Rev. Lett. \textbf{101} (2008) 042001.

\bibitem{ln_hermes}
A.~Airapetian \textit{et al.}, Phys. Rev. Lett. \textbf{103} (2009) 152002;
Phys. Lett. \textbf{B693} (2010)~11.

\bibitem{ln_compass}
M.G.~Alekseev \textit{et al.}, Phys. Lett. \textbf{B692} (2010) 240.

\bibitem{ln_sivers}
D.~Sivers, Phys. Rev. \textbf{D41} (1990) 83; \textbf{D43} (1991) 261.

\bibitem{ln_collins}
J.C.~Collins, Phys. Lett. \textbf{B536} (2002) 43.

\bibitem{ln_kang}
Z.~Kang, J.~Qiu, W.~Vogelsang, F.~Yuan, Phys. Rev. \textbf{D83} (2011) 094001.

\bibitem{ln_ap}
A.~Prokudin, Z.~Kang, workshop ''Opportunities for Drell-Yan Physics at RHIC'',\\
http://www.phenix.bnl.gov/WWW/publish/elke/TALKS/DY-May2011/Talks/AlexeiProkudin.pdf.


\bibitem{ln_cp_dis2011}
C.~Perkins, arXiv:1109.0650.

\end{thebibliography}
\end{document}